\documentclass{article}
\usepackage{spconf,amsmath,graphicx,hyperref}
\usepackage{booktabs}
\usepackage{xcolor}
\usepackage{microtype}
\usepackage{xspace}
\usepackage{flushend}

\hypersetup{hidelinks}

\newcommand{\model}{CPR\xspace}

\title{CPR: COMBINING GLOBAL COMPOSING, LOCAL PERFORMING AND FULL-SEQUENCE REFINING IN PIANO RENDERING WITH CONTINUOUS AUTOREGRESSIVE MODELLING}
\name{Chong Jing, Junan Zhang, Zhizheng Wu\thanks{\normalsize Corresponding author: Zhizheng Wu.}}
\address{The Chinese University of Hong Kong, Shenzhen\\
{\small\texttt{\{chongjing,junanzhang\}@link.cuhk.edu.cn, wuzhizheng@cuhk.edu.cn}}}

\begin{document}
\ninept
\maketitle

\begin{abstract}
Prompt-conditioned piano MIDI-to-Music rendering aims to faithfully render target notes while reproducing the timbre of a reference recording. Existing approaches primarily follow two paradigms: autoregressive (AR) modeling and flow matching (or diffusion). Discrete-codec AR models provide causal temporal modeling, but quantization can discard acoustic detail. Flow matching better preserves acoustic structure at the cost of full-sequence attention and weaker semantic structure. Continuous autoregressive models operate directly on continuous representations. They not only combine the condition-following ability of AR models with the distribution-modeling capacity of flow matching but also bypass the quantization bottleneck with lower computational costs. Building on this principle, we present the \textbf{C}omposer--\textbf{P}erformer--\textbf{R}efiner (\textbf{\model}) framework. Composer autoregressively predicts continuous hidden states, Performer generates 24 kHz acoustic latents through local flow matching, and Refiner then upsamples the waveform to 48 kHz. We further introduce Bottlenecked Representation Alignment (\textbf{BREPA}) and Modality--Time RoPE (\textbf{MT-RoPE}) to strengthen musical semantic structure in Composer hidden states and temporal alignment across modalities. Code is available at \url{https://github.com/FEAfeatherTHER/CPR_official}.
\end{abstract}


\section{Introduction}
\label{sec:intro}
Prompt-conditioned piano MIDI-to-Music synthesis renders target MIDI while transferring the acoustic identity of a reference recording. Given prompt audio, its aligned MIDI, and target MIDI, a piano renderer must follow note onsets and dynamics, preserve prompt timbre, and produce a musically expressive performance.

Autoregressive (AR) modeling and flow matching (or diffusion)~\cite{lipman2023flow,song2021scorebasedgenerativemodelingstochastic} are two established paradigms for systems like MIDI-to-Music (MTM) and Text-to-Speech (TTS)~\cite{tang2025midivalle,hu2026qwen3ttstechnicalreport,copet2024simplecontrollablemusicgeneration,jing2026pmuse,jing2026anysynthzeroshotinstrumentcloningincontext,chen2025f5ttsfairytalerfakesfluent}. AR models generate audio through next-token prediction, as in MIDI-VALLE~\cite{tang2025midivalle}. However, quantization limits the representation of continuous music, while the train--test mismatch from teacher forcing causes cumulative errors, degrading long-sequence generation and controllability.
Full-sequence flow matching with bidirectional attention can capture richer acoustic details, as demonstrated by P-MUSE~\cite{jing2026pmuse}. It demonstrates superior timbre similarity but lacks naturalness and musicality. Prior studies~\cite{yu2024repa,xu2026ifid} have shown that pure flow matching supervision struggles to recover the semantic structure necessary for generation and therefore requires auxiliary alignment objectives.

A common improved hybrid system uses an AR model to predict first-depth codebook tokens that condition full-sequence flow matching~\cite{zhang2025minimaxspeechintrinsiczeroshottexttospeech}. These independently trained stages combine both paradigms, but the quantization bottleneck still limits generation performance. Several recent works have proposed autoregressively generating continuous representations to bypass the quantization bottleneck. DiTAR~\cite{jia2025ditar} builds a continuous autoregressive TTS system composed of a global autoregressive model and a local diffusion model with aggregated patch tokens, which shorten sequences and thus mitigate cumulative error. FireRedTTS-3~\cite{shen2026fireredtts3} further emphasizes the semantic structure of latents because of its importance in continuous autoregressive models and the trade-off between reconstruction and generation~\cite{xu2026ifid}.

Considering the more complex temporal structure and harmonic components in the music domain, MiniMax Music 3~\cite{minimax2026music3} designs a cascaded system with a Hybrid-LLM and a full-sequence flow matching transformer. Hybrid-LLM incorporates a Global-LLM and a Local-LLM to regress codes of the first and residual depths, respectively. To mitigate quantization loss, its generation is directly conditioned on continuous Hybrid-LLM hidden states.
However, the Local-LLM can worsen error accumulation, so replacing it with local FM is more fundamentally aligned with first principles to alleviate both limitations. It is also worth noting that the tokenizer in MiniMax Music 3 operates at 24 kHz, while its variational autoencoder (VAE) operates at 44.1 kHz. This strategy arises from the different representational requirements and preferences of autoregressive modeling and flow matching.

Building on these observations, we propose the \textbf{C}omposer--\textbf{P}erformer--\textbf{R}efiner (\model) framework for MIDI-to-Music. We first jointly train Composer--Performer as a continuous autoregressive model on 24 kHz audio. We then train a Refiner~\cite{siuzdak2023vocos} to upsample the generated waveforms to 48 kHz. Since the flow matching loss lacks direct supervision for Composer hidden states, we introduce Bottlenecked Representation Alignment (BREPA) as an auxiliary semantic supervision objective. It aligns a compressed projection of Composer hidden states with self-supervised features, strengthening musical structure while retaining capacity for acoustic details. To further enhance condition following, we introduce Modality--Time Rotary Position Embeddings (MT-RoPE) to improve cross-modal temporal alignment between MIDI and audio.

\begin{figure*}[t]
\centering
\includegraphics[width=\textwidth]{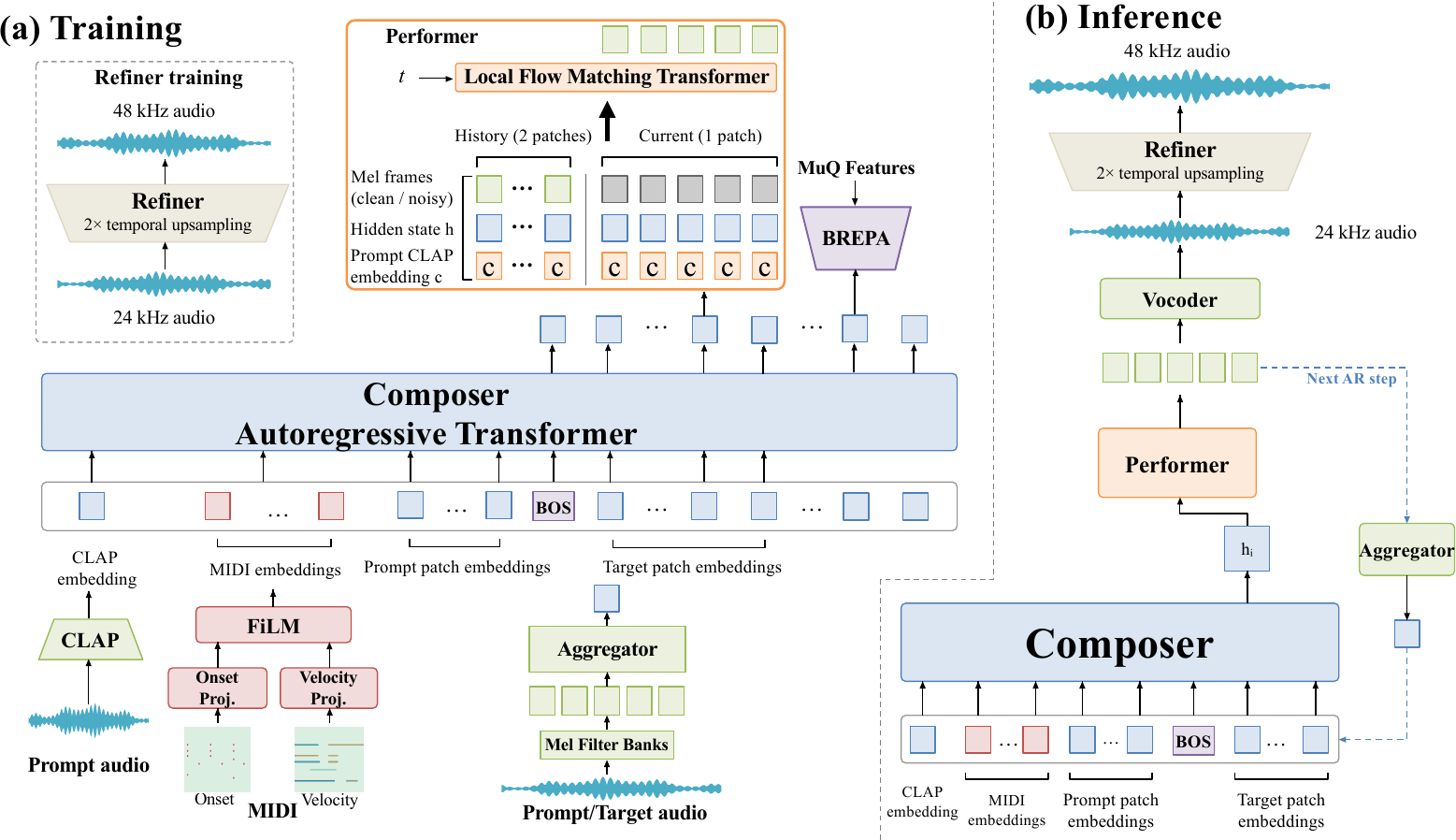}
\caption{Overview of \model: (a) training, including the Performer architecture, and (b) autoregressive inference. Composer and Performer are jointly trained with flow matching and BREPA losses. Performer conditions each 5-frame patch on two clean historical patches, concatenating Composer hidden states and repeated prompt CLAP embeddings with mel frames along the feature dimension. The separately trained Vocos-based Refiner maps generated 24 kHz audio to 48 kHz.}
\label{fig:cpr}
\end{figure*}

In summary, the contributions are as follows: 
\begin{itemize}
    \item We introduce \model, a framework that combines complementary strengths of autoregressive and flow matching modeling. Extensive experiments demonstrate that \model\ achieves state-of-the-art (SOTA) performance on the piano MIDI-to-Music generation task.
    \item To enhance semantic alignment and facilitate the joint optimization of the system, we introduce Bottlenecked Representation Alignment (BREPA).
    \item We propose Modality--Time RoPE (MT-RoPE), a two-axis adaptation of rotary position embeddings~\cite{su2023roformerenhancedtransformerrotary} that achieves better temporal alignment between audio and MIDI modalities.
\end{itemize}

\section{Method}
\label{sec:method}

\subsection{Formulation}
\label{sec:formulation}

Given a sequence of continuous embeddings ($x_1, x_2, \ldots, x_N$), an autoregressive model performing next-token prediction can be formulated as:
\begin{equation}
p_\theta(x_1, x_2 ,\ldots, x_N)=\prod_{i=1}^{N}p_\theta(x_i \mid x_1,x_2,\ldots,x_{i-1}).
\label{eq:ar}
\end{equation}

Instead of regressing token by token, \model aggregates groups of $P$ adjacent tokens into one patch embedding $p_i$ and replaces the classification head based on discrete tokens with a local flow matching head based on continuous latents. Local flow matching models the distribution $p_\theta(x_{i+1}, \ldots, x_{i+P})$ of the current patch.

\subsection{Architecture}
\label{sec:architecture}

\subsubsection{Composer--Performer--Refiner}

As shown in Fig.~\ref{fig:cpr}, \model is trained in two stages. 
The first stage jointly trains a continuous autoregressive system comprising Composer and Performer.
We aggregate five 20-ms Mel frames into one 100-ms patch embedding.
Composer is a causal transformer that autoregressively predicts the continuous hidden states of each acoustic patch, while Performer is a local bidirectional flow matching transformer~\cite{peebles2023scalablediffusionmodelstransformers} that generates the next acoustic patch conditioned on the CLAP embedding, Composer hidden states, and historical information through in-context learning.
The second stage trains the Vocos-based Refiner~\cite{siuzdak2023vocos}, which efficiently upsamples the generated 24 kHz waveform to 48 kHz.

\subsubsection{Composer}

Prompt and target MIDI are rasterized at 25 Hz into onset and velocity channels. Separate projections encode the two channels, which are fused through feature-wise linear modulation (FiLM) to form MIDI embeddings. The patch embeddings are aggregated into a prepended \texttt{[CLS]} token through a transformer encoder within each patch. The prompt audio's CLAP embeddings~\cite{wu2024largescalecontrastivelanguageaudiopretraining}, MIDI embeddings, and patch embeddings are concatenated along the sequence dimension. The backbone of Composer is an autoregressive transformer.

To strengthen cross-modal temporal alignment, we introduce Modality--Time RoPE (MT-RoPE). It is a two-axis adaptation of rotary position embeddings~\cite{su2023roformerenhancedtransformerrotary}. Inspired by the axis-wise decomposition in Qwen2-VL's M-RoPE~\cite{wang2024qwen2vl} and ARDiT's modality-specific fractional position indexing~\cite{liu2024autoregressivediffusiontransformertexttospeech}, we allocate half of the rotary dimensions to modality identity and half to temporal position. MIDI and audio embeddings are indexed independently. For MIDI frame $r$ at 25 Hz and audio patch $i$ at 10 Hz, their temporal coordinates are $\frac{10}{25}r$ and $i$, respectively, placing both modalities on a common physical-time axis.

\subsubsection{Performer}
Flow matching models the data distribution through a velocity field conditioned on the noisy observation. We use a linear path~\cite{lipman2023flow} between clean data and standard Gaussian noise:
\begin{equation}
    x_t = t x_1 + (1-t) x_0, \quad \text{where } x_0 \sim \mathcal{N}(0, \mathbf{I})
\end{equation}
Along the sequence dimension, we concatenate the historical clean context with the current noisy mel spectrogram. Along the feature dimension, we concatenate the audio prompt's CLAP embedding, Composer hidden states, and mel spectrogram. CLAP supplies a global timbre prior when local history is degraded or uninformative. To prevent this prior from becoming a shortcut that limits in-context learning, Performer uses hierarchical condition dropping for CFG~\cite{ho2022classifierfreediffusionguidance}: dropping Composer hidden states always drops CLAP; otherwise, CLAP is dropped with probability 0.2. This encourages timbre cloning through history and Composer hidden states.

\subsubsection{Refiner}
Because patches aggregated from latents of 48 kHz audio contain too much high-frequency acoustic information unsuitable for semantic modeling, our continuous autoregressive model is trained on 24 kHz audio.
Then Refiner, composed of a transposed convolution layer, a ConvNeXt backbone, and an STFT prediction head~\cite{siuzdak2023vocos}, converts the reconstructed 24 kHz waveform to 48 kHz to compensate for missing high-frequency harmonic components.

\subsubsection{Bottlenecked Representation Alignment}

BREPA operates only on the alignment branch during training~\cite{zhu2025muq}. A transposed-convolution module upsamples Composer hidden states from 10 Hz to 25 Hz with a bottleneck ($1024 \rightarrow 768$) enforcing irreversible compression. The resulting representations are aligned with MuQ features using a cosine similarity loss. The bottleneck forces most dimensions of the hidden states to capture semantic structure while the remaining dimensions retain capacity for residual information like acoustic details.

\section{Experiments}
\label{sec:experiments}

\subsection{Dataset}
\label{sec:dataset}

The training data combine real recordings and synthesized audio with aligned MIDI annotations, restricted to piano-family instruments. They comprise real piano performances from MAESTRO~\cite{hawthorne2019maestro}, piano tracks from Slakh~\cite{manilow2019slakh}, and single-track MIDI sequences from Lakh~\cite{raffel2016lakh} rendered using NSynth note samples~\cite{engel2017nsynth}. The overall timbral distribution basically covers all eight subcategories within the Piano family under the General MIDI standard. Recordings are segmented into 3--30-s clips, totaling approximately 5,000 hours in duration. We evaluate on the paired-prompt piano generation task of P-MUSE-eval~\cite{jing2026pmuse}, comprising 100 samples covering diverse piano-family timbres. Each sample provides an audio prompt, its aligned MIDI, and a target MIDI sequence.

\subsection{Implementation Details}
\label{sec:implementation}

\noindent\textbf{Model.}
Composer is initialized from Qwen3-0.6B~\cite{yang2025qwen3} with the original vocabulary embeddings removed. The patch aggregator is a 4-block transformer encoder with 8 attention heads, a hidden dimension of 1024, and a feed-forward dimension of 4096. Composer contains approximately 500M parameters.
Performer is a 6-block DiT with 8 attention heads, a hidden dimension of 1024, and a feed-forward dimension of 4096, totaling approximately 175M parameters. We extract 128-dimensional mel spectrograms from 24 kHz audio with a hop size of 480 samples.
The vocoder is based on Vocos~\cite{siuzdak2023vocos} and maps generated mel spectrograms to 24 kHz waveforms, with approximately 255M parameters. 
Refiner is also Vocos-based with a transposed-convolution module and contains approximately 14M parameters.

\noindent\textbf{Training.}
Composer and Performer are trained jointly using AdamW~\cite{loshchilov2019decoupledweightdecayregularization} with $\beta_1=0.9$, $\beta_2=0.999$, weight decay 0.01, and a gradient clipping threshold of 0.2. Training is performed on 8 NVIDIA GeForce RTX 5090 GPUs with dynamic batch sizes for 200,000 steps, with a 4,000-step linear warmup followed by inverse-square-root learning-rate decay. The peak learning rate is $2\times10^{-5}$ for the Qwen3-initialized AR Transformer and $1\times10^{-4}$ for the remaining modules in the Composer--Performer system. The training objective for Composer--Performer is
\begin{equation}
\mathcal{L}_{\mathrm{C-P}} = \mathcal{L}_{\mathrm{flow}} + 0.5\,\mathcal{L}_{\mathrm{BREPA}}.
\label{eq:cp_loss}
\end{equation}

Refiner is pretrained for 700k steps and then fine-tuned for 90k steps on full-band data. Random bandwidth degradation constructs inputs paired with the original 48 kHz audio as targets. We use AdamW with a batch size of 48. Generator and discriminator learning rates are $10^{-4}$ during pretraining. During fine-tuning, the generator rate is reduced to $2\times10^{-5}$ while the discriminator rate remains $10^{-4}$.
The Refiner generator objective is
\begin{equation}
\begin{aligned}
\mathcal{L}_{\mathrm{Refiner}}
&= \mathcal{L}_{\mathrm{MR\text{-}STFT}} + 15\,\mathcal{L}_{\mathrm{MR\text{-}Mel}} + \mathcal{L}_{\mathrm{adv}} + \mathcal{L}_{\mathrm{feat}},
\end{aligned}
\label{eq:refiner_loss}
\end{equation}
where the four terms denote multi-resolution STFT, multi-resolution mel, adversarial, and feature-matching losses, respectively~\cite{kong2020hifigangenerativeadversarialnetworks}.

\noindent\textbf{Inference.}
Each patch is sampled from Gaussian noise using a first-order Euler ODE solver. Let $v_c=v_\theta(x_t,t,e,h,H)$ denote the conditional velocity field, where $e$ is the audio prompt's CLAP embedding, $h$ is the Composer hidden state, and $H$ is the clean historical context. Dropping only $e$ and $h$ while retaining $H$ gives $v_u=v_\theta(x_t,t,\emptyset,\emptyset,H)$. CFG and Euler updates are
\begin{align}
v_{\mathrm{CFG}} &= v_c+w(v_c-v_u)
\label{eq:cfg}\\
x_{t_{k+1}} &= x_{t_k}+(t_{k+1}-t_k)v_{\mathrm{CFG}}.
\label{eq:euler}
\end{align}
Here, $w$ is the CFG strength. We use 4 Euler integration steps per patch. After sampling, the generated mel patch is converted into a patch embedding and fed back to Composer for the next autoregressive step. The generated mel sequence is decoded into 24 kHz audio by the vocoder and mapped to 48 kHz by Refiner.

\begin{table}[!ht]
\caption{Objective results on P-MUSE-eval test set. C--P: outputs of Composer--Performer; SR: sampling rate in kHz. Sim and onset F1 are percentages. NFE counts sampling steps; Refiner adds none. ``--'' means not applicable. Best and second-best metric values are bold and underlined.}
\label{tab:main}
\centering
\normalsize
\setlength{\tabcolsep}{2pt}
\begin{tabular}{@{}lrrrrr@{}}
\toprule
Model & SR & NFE & Sim $\uparrow$ & Onset F1 $\uparrow$ & $\mathrm{FAD}_{\mathrm{clap}}\downarrow$ \\
\midrule
P-MUSE & 24 & 25 & \textbf{90.6} & 75.7 & 0.161 \\
MIDI-VALLE & 32 & -- & 83.4 & 63.7 & 0.390 \\
\midrule
C--P & 24 & 2 & 89.0 & 75.1 & 0.165 \\
C--P & 24 & 4 & \underline{89.3} & \underline{78.3} & \underline{0.147} \\
C--P & 24 & 10 & \underline{89.3} & \textbf{78.4} & 0.148 \\
\model & 48 & 4 & \underline{89.3} & \textbf{78.4} & \textbf{0.129} \\
\bottomrule
\end{tabular}
\end{table}

\begin{table}[!ht]
\caption{MOS on 14 P-MUSE-eval samples rated by 13 listeners on a 1--5 scale. Higher is better.}
\label{tab:mos}
\centering
\normalsize
\setlength{\tabcolsep}{3pt}
\begin{tabular}{@{}lrrr@{}}
\toprule
Model & Timbre sim. & MIDI acc. & Naturalness \\
\midrule
Ground truth & 4.429 & 4.929 & 4.500 \\
\midrule
P-MUSE & \underline{4.071} & 4.571 & 3.857 \\
MIDI-VALLE & 2.357 & 3.643 & 3.214 \\
\midrule
C--P (24 kHz) & \textbf{4.214} & \underline{4.643} & \underline{4.000} \\
\model (48 kHz) & 4.000 & \textbf{4.714} & \textbf{4.071} \\
\bottomrule
\end{tabular}
\end{table}

\subsection{Metrics}

We report timbre similarity (Sim), onset F1, and Fr\'{e}chet Audio Distance (FAD)~\cite{kilgour2018fad} to evaluate timbre preservation, MIDI adherence, and overall generation quality, respectively.

Sim measures the embedding similarity between generated audio and its corresponding audio prompt~\cite{shi2021usespeakerrecognitionapproaches}. For onset F1, we use \texttt{MuScriptor-Large}~\cite{rouard2026muscriptor} to transcribe the generated audio into note sequences, which are then compared with the target MIDI. A note is considered correctly matched only if its pitch matches the reference and its onset is within 50 ms of the reference onset (offsets are not considered). For FAD, we use LAION-CLAP (music) features~\cite{wu2024largescalecontrastivelanguageaudiopretraining} to compute the Fr\'{e}chet distance between the distributions of the generated audio set and the corresponding ground-truth target audio set from P-MUSE-eval, denoted as $\mathrm{FAD}_{\mathrm{clap}}$. Higher Sim and onset F1 and lower $\mathrm{FAD}_{\mathrm{clap}}$ indicate better performance.

For subjective evaluation, we report mean opinion scores (MOS) for timbre similarity, MIDI accuracy, and naturalness. Listeners rate each dimension on a 1--5 scale with higher scores indicating better performance. We extract 14 samples from the P-MUSE-eval test set, covering all timbres with one sample per timbre. Thirteen listeners participate in a fully blind evaluation. Ground-truth audio is also included for comparison.

\subsection{Results}

\noindent\textbf{Objective evaluation.} We compare against P-MUSE~\cite{jing2026pmuse} and MIDI-VALLE~\cite{tang2025midivalle}. P-MUSE is a single-stage non-autoregressive generator based on a full-sequence flow matching Transformer. MIDI-VALLE uses a discrete audio tokenizer and a two-stage generation pipeline: an AR model predicts the first codebook, followed by an NAR model that generates the remaining codebooks. Since MIDI-VALLE accepts only a 3-second audio prompt, we use the first 3 seconds of the original prompt for this baseline.

As shown in Table~\ref{tab:main}, \model outperforms MIDI-VALLE on all three evaluation metrics. Compared with P-MUSE, \model achieves slightly lower timbre similarity. One possible explanation is that P-MUSE's full-sequence flow matching Transformer can jointly exploit broader context to capture richer timbral details. In contrast, \model improves onset F1 by 2.7 percentage points, indicating more accurate adherence to target MIDI pitches and onsets and better cross-modal temporal alignment. For overall generation quality, \model achieves the lowest $\mathrm{FAD}_{\mathrm{clap}}$. These results indicate that the generated audio more closely matches the real-audio distribution, providing distribution-level support for improved overall musicality and naturalness.

Table~\ref{tab:main} also demonstrates the metric trends with respect to the number of function evaluations (NFE). The performance peaks near $\text{NFE} = 4$, beyond which further increases in NFE bring no additional benefit. Unless stated otherwise, \model operates with $\text{NFE} = 4$ for all remaining evaluations.

\noindent\textbf{Subjective evaluation.}
We select 14 samples from the P-MUSE-eval test set, covering all timbres with one sample per timbre. Thirteen listeners participate in a fully blind evaluation. Overall, \model's subjective evaluation results (Table~\ref{tab:mos}) are consistent with the objective metrics. Introducing Refiner increases the scores for MIDI accuracy and for naturalness and musicality, but reduces timbre similarity, revealing a trade-off across perceptual dimensions. This suggests limitations of a purely mapping-based super-resolution design for Refiner, which is unable to restore detailed information lost during low-resolution generation. Designing a Refiner that effectively recovers such information is left for future work.

\begin{table}[!ht]
\caption{Controlled ablations on P-MUSE-eval. Each variant changes only the indicated setting; all other settings remain fixed. Sim and onset F1 are percentages.}
\label{tab:ablation}
\centering
\normalsize
\setlength{\tabcolsep}{2pt}
\begin{tabular}{@{}lrrr@{}}
\toprule
Model & Sim $\uparrow$ & Onset F1 $\uparrow$ & $\mathrm{FAD}_{\mathrm{clap}}\downarrow$ \\
\midrule
\model & \textbf{89.3} & \textbf{78.4} & \textbf{0.129} \\
w/o REPA & \underline{89.1} & 73.2 & 0.145 \\
REPA w/o Bottleneck & 88.7 & \underline{78.1} & \underline{0.136} \\
standard RoPE & 88.0 & 64.4 & 0.251 \\
\bottomrule
\end{tabular}
\end{table}

\subsection{Ablations}

We conduct three controlled ablations to examine BREPA and MT-RoPE. Each experiment changes only the module under study; all other model components, training data, training configurations, inference settings, and evaluation procedures remain unchanged. The three variants remove REPA, use REPA without the bottleneck, or replace MT-RoPE with standard one-dimensional RoPE, respectively.

As shown in Table~\ref{tab:ablation}, training with REPA can improve onset F1 and $\mathrm{FAD}_{\mathrm{clap}}$, and introducing the bottleneck brings additional improvements. This indicates that introducing the REPA objective into training reshapes the manifold of Composer hidden states, encouraging them to retain more rhythmic, pitch, and dynamic structure, thereby improving cross-modal control and overall generation quality. The inner bottleneck can further strike a balance between acoustic and semantic structures in Composer hidden states. Replacing MT-RoPE with standard RoPE causes a larger degradation despite retaining BREPA during training, indicating that semantic representation constraints cannot fully replace explicit cross-modal temporal alignment.

\section{Conclusion}
\label{sec:conclusion}

We introduced \model, combining global autoregressive modeling, local flow matching, and full-sequence refinement.
BREPA and MT-RoPE enhance semantic and cross-modal temporal alignment.
Future work may include using low-resolution observations for efficient refinement at higher sampling rates, exploring representations for continuous autoregressive music modeling, and incorporating richer conditioning for instruction following.

\section{Compliance with Ethical Standards}
\label{sec:ethics}

All listeners voluntarily participated in the subjective listening evaluation with informed consent, and their ratings were collected anonymously.

\section{Acknowledgments}
\label{sec:acknowledgments}

The authors declare no financial or nonfinancial conflicts of interest related to this work.

\bibliographystyle{IEEEbib}
\bibliography{references}

\end{document}